\documentclass{article}
\usepackage{graphicx}
\usepackage{amsmath}
\usepackage{amsfonts}
\usepackage{amssymb}
\begin{document}

\title{\textbf{DUALITIES IN FRACTIONAL STATISTICS}}
\author{{\normalsize M. I. BECIU}\\{\footnotesize Department of Theoretical Physics, Institute of Atomic Physics}\\{\footnotesize Magurele, Sect 6, P.O. Box MG-6, Bucharest, Romania}\\{\footnotesize Department of Physics, Technical University B-d Lacul Tei 124,
Bucharest, Romania}\\{\footnotesize beciu@mail.utcb.ro}}
\maketitle
\begin{abstract}
{\footnotesize We first reobtain in a simpler way the Haldane fractional
statistics at thermal equilibrium using an interpolation argument. We then
show that the mean occupation number for fractional statistics is invariant to
a group of duality transformations, a nonabelian subgroup of fractional linear group.}
\end{abstract}

\section{\textbf{Introduction}}

\noindent{\normalsize In the two-dimensional systems the particles are not
constrained to obey the Fermi or Bose statistics but they may have fractional
statistics; two well known exemples are offered by the quasi-particles in the
fractional quantum Hall effect and by anyons. Prompted by this, Haldane
\cite{1} stated a generalized exclusion principle that allows for intermediate
statistics, though his derivation of the particle occupation number does not
involve the spatial dimensionality and so, it seems to be valid for an
arbitrary number of dimensions. Wu \cite{2} discussed the thermodynamics of
the fractional statistics particles (geons) in the ideal gas approximation.
Further developments on the thermodynamical aspects of geons can be found in
\cite{3}, \cite{4}. In an akin paper\cite{5} to the present one, it was shown
that the distribution funtion for geons enjoys a duality property. Quite
recently, a number of authors \cite{6}, starting from an appropiate definition
for indistinguishability in the phase space, have obtained a nontrivial
classical counterpart of Haldane's exclusion principle.}

{\normalsize The purpose of this Letter is first, to rederive, using a very
simple interpolating argument, the occupation number distribution for geons at
thermal equilibrium. Then, based on this derivation, we show that the relation
for the occupation number remains invariant under a certain number of dual
transformations that organize in a group and we discuss the meaning of these dualities.}

\bigskip

\section{Model}

{\normalsize A} {\normalsize very long time ago, when the black body radiation
formula was still under scrutiny, a simple argument based on interpolation has
been put forward in \cite{7}. The argument goes as follows: Let be }%
$\overline{\varepsilon}$ {\normalsize the mean energy for a radiation field,
averaged over time in a fixed unit volume. Using the equality of time averages
and phase spase averages one can write for the mean squares values }%
$\overline{\Delta\varepsilon^{2}}${\normalsize \ of the fluctuating energy that}%

\begin{equation}
\overline{\Delta\varepsilon^{2}}=-k\left(  \frac{\partial\overline
{\varepsilon}}{\partial\left(  \frac{1}{T}\right)  }\right)  _{V}
\index{1} \label{this}%
\end{equation}

{\normalsize Empirically, the mean square value gets contributions from low
frequencies (the Rayleigh-Jeans part of the spectrum), where the mean energy
}$\overline{\varepsilon}=-kT${\normalsize ; therefore}%

\begin{equation}
kT^{2}\frac{\partial\overline{\varepsilon}}{\partial T}=\overline{\varepsilon
}^{2}%
\end{equation}
{\normalsize It also gets contributions from high frequencies}
{\normalsize (the Wien part of the spectrum), where the mean energy
}$\overline{\varepsilon}=h\nu\exp\left(  -\frac{h\nu}{kT}\right)
${\normalsize ; hence}%

\begin{equation}
kT^{2}\frac{\partial\overline{\varepsilon}}{\partial T}=h\nu\overline
{\varepsilon}%
\end{equation}

{\normalsize With the underlying assumption that the two contributions act
independently and are of equal weight, relation (1) reads as}%

\begin{equation}
kT^{2}\frac{\partial\overline{\varepsilon}}{\partial T}=\overline{\varepsilon
}h\nu+\overline{\varepsilon}^{2}%
\end{equation}

{\normalsize It is more convenient to use the dimensionless variables
}$n=\dfrac{\overline{\varepsilon}}{\varepsilon}${\normalsize , }$\dfrac{h\nu
}{kT}=\alpha$ {\normalsize that lead to the differential equation:}%

\begin{equation}
-\frac{dn}{d\alpha}=n+n^{2}%
\end{equation}

{\normalsize The solution is}%

\begin{equation}
n=\frac{1}{\exp\left(  \alpha-\alpha_{0}\right)  -1}\text{ or }\overline
{\varepsilon}=\frac{\varepsilon}{\exp\left(  \dfrac{\varepsilon-\mu}%
{kT}\right)  -1}%
\end{equation}

{\normalsize This is, the Planck law, here with }$\alpha=\dfrac{\mu}{kT}%
${\normalsize , merely an integration constant. This derivation emphasises the
wave-particle character of bosons.}

{\normalsize How this reasoning can be extended for other statistics? Note
that this type of interpolation works equally well for fermions. However, for
fermions, as long as }$\varepsilon<\mu${\normalsize , the long wavelength
(high temperature) is negative:}%

\begin{equation}
\frac{d\varepsilon}{dT}\sim-\overline{\varepsilon}^{2}%
\end{equation}
{\normalsize and accordingly, relation (5) must be modified to}%

\begin{equation}
-\frac{dn}{d\alpha}=n-n^{2}%
\end{equation}
{\normalsize When we take into account that }$n\leq1${\normalsize , we obtain,
as solution, the Fermi distribution:}%

\begin{equation}
n=\frac{1}{\exp\left(  \alpha-\alpha_{0}\right)  +1}%
\end{equation}

{\normalsize Introducing a genuine interpolation parameter \textit{g}, i.e. a
parameter ranging in the }$\left[  0,1\right]  ${\normalsize \ interval, one
can write in a single stroke, for both fermions and bosons:}%

\begin{equation}
-\frac{dn}{d\alpha}=n+\left(  1-2g\right)  n^{2}%
\end{equation}
{\normalsize with the solution}%

\begin{equation}
n=\frac{1}{\exp\left(  \alpha-\alpha_{0}\right)  -\left(  1-2g\right)  }%
\end{equation}

{\normalsize Note that this represents also, as a particular case, for
}$g=\frac{1}{2}${\normalsize , the Maxwell distribution function. The above
interpolating formula has also been obtained in Refs.8.}

{\normalsize At this moment we may ask what other modifications must be
brought to (9) if one would allow for intermediate statistics betweeen Bose
and Fermi ones on one hand, relation (9) suggests a series expension in powers
of mean occupation number \textit{n}. This is what we might expect because an
interpolating statistics presumely contains more parameters and its complete
characterization needs higher moments and/or higher powers of the first
moment. On the other hand, higher order terms in (9) must be multiplied, for
sure, by the factor }$g\left(  g-1\right)  ${\normalsize , that excludes for
standard fermionic }$\left(  g=1\right)  ${\normalsize \ and bosonic }$\left(
g=0\right)  ${\normalsize \ statistics any extra terms which, obviously, do
not occur in (10). Limiting ourselves to the next higher order term }$n^{3}%
${\normalsize \ and using again the independence and equal weights assumption
(with the exception of }$g\left(  g-1\right)  ${\normalsize \ factor), we
arrive at}%

\begin{equation}
-\frac{dn}{d\alpha}=n+\left(  1-2g\right)  n^{2}+g\left(  g-1\right)  n^{3}%
\end{equation}

{\normalsize The differential equation can be solved by an elementary
integration. To this purpose and for later reference we make the change of variable:}%

\begin{equation}
w=\frac{1-ng}{n}%
\end{equation}

{\normalsize The differential equation transforms to}%

\begin{equation}
\frac{dw}{d\alpha}=\frac{w\left(  w+1\right)  }{w+g}%
\end{equation}

{\normalsize The solution is}%

\begin{equation}
w^{g}\left(  w+1\right)  ^{1-g}=e^{\left(  \alpha-\alpha_{0}\right)  }%
\end{equation}

{\normalsize Formulas (13) and (15) together gives implicitely the mean
occupation number as function of inverse temperature. The physical meaning of
the parameter \textit{g} can be drawn from the zero temperature limit
(}$\alpha\longrightarrow\infty):$%

\begin{equation}
n=\left\{
\begin{array}
[c]{cc}%
0 & \varepsilon>\varepsilon_{F}\\
\frac{1}{g} & \varepsilon<\varepsilon_{F}%
\end{array}
\right.
\end{equation}

{\normalsize Therefore, }$\frac{1}{g}${\normalsize \ represents the maximum
number of particles a quatum state can accommodate. Actually, formulas (13),
(15) represent exactly the particle distribution for geons at thermal
equilibrium obtained by Haldane \cite{1} from a counting argument and from
equilibrium condition requirements while (16) expresses the generalized
exclusion principle.}

{\normalsize We have unfolded the derivation starting from the mean square
value of the energy just to remind an historical argument \cite{7}. Of course
we could have started from the mean square value of the mean number directly.
It is an well-known fact that the fluctuations in the number of bosons is super-poissonian:}%

\begin{equation}
\overline{\Delta n^{2}}=\overline{n}+\overline{n}^{2}%
\end{equation}
{\normalsize while for fermions is sub-poissonian:}%

\begin{equation}
\overline{\Delta n^{2}}=\overline{n}-\overline{n}^{2}%
\end{equation}

{\normalsize The advantage of the above derivation of fractional statistics
over the standard one consists in its simplicity. The shortcoming lies in the
fact the formulas for \textit{n} are valid only at equilibrium.}

{\normalsize What are the transformations that leave invariant the occupation
number formulas invariant, i.e. formulas (13) and (15)? To be more specific,
we look for transformations for }$w${\normalsize \ that, while preserving the
substitution (13) intact, makes relation (15), or, under the same boundary
condition, the differantial equation (14), form-invariant. An earlier
derivation for such a tranformation \cite{5} and a theoretical prejudice lead
us to consider transformations of the form:}%

\begin{equation}
\widetilde{w}=\frac{aw+b}{cw+d}%
\end{equation}

{\normalsize They are known as transformations belonging to the modular group
SL(2,G), or the fractional linear group, where G stands for C, R or Z
depending on whether the coefficients \textit{a}, \textit{b}, \textit{c},
\textit{d} are complex, real or integer, respectively. The group
multiplication law may be represented by a 2}$\times${\normalsize 2 matrix
}$\gamma=\left(
\begin{array}
[c]{cc}%
a & b\\
c & d
\end{array}
\right)  ${\normalsize \ with unit determinant \textit{ad}-\textit{cd}=1. The
theoretical prejudice alluded above refers to the fact the SL(2,Z) has been
found relevant for the complex conductivities in the quantum Hall effect
\cite{8}.}

{\normalsize We substitute Eq. (19) into the differential equation (14) and
look for the coefficients \textit{a, b, c} and \textit{d} that leave the
equation in same form}.{\normalsize \ Rather long but straightforward algebra
shows that, excepting the identity, there are only five possible
tranformations of type (19) that leave the geon distribution form-invariant.
These transformations }$\left(  t_{k}\right)  ${\normalsize \ are:}%

\[%
\begin{tabular}
[c]{c||c|c|c}%
$%
%TCIMACRO{\QDATOP{\QTATOP{}{}}{\QTATOP{}{}}}%
%BeginExpansion
\genfrac{}{}{0pt}{0}{\genfrac{}{}{0pt}{1}{}{}}{\genfrac{}{}{0pt}{1}{}{}}%
%EndExpansion%
%TCIMACRO{\QDATOP{\QTATOP{}{}}{\QTATOP{}{}}}%
%BeginExpansion
\genfrac{}{}{0pt}{0}{\genfrac{}{}{0pt}{1}{}{}}{\genfrac{}{}{0pt}{1}{}{}}%
%EndExpansion
$ & $\hspace{1cm}\widetilde{w}\hspace{1cm}$ & $\hspace{1cm}\widetilde
{g}\hspace{1cm}$ & $\hspace{1cm}\widetilde{\alpha}\hspace{1cm}$\\\hline\hline
$%
%TCIMACRO{\QDATOP{\QDATOP{}{}}{\QDATOP{}{}}}%
%BeginExpansion
\genfrac{}{}{0pt}{0}{\genfrac{}{}{0pt}{0}{}{}}{\genfrac{}{}{0pt}{0}{}{}}%
%EndExpansion
t_{1}%
%TCIMACRO{\QDATOP{\QDATOP{}{}}{\QDATOP{}{}}}%
%BeginExpansion
\genfrac{}{}{0pt}{0}{\genfrac{}{}{0pt}{0}{}{}}{\genfrac{}{}{0pt}{0}{}{}}%
%EndExpansion
$ & $\dfrac{1}{w}$ & $\dfrac{1}{g}$ & $-\dfrac{\alpha}{g}$\\\hline
$%
%TCIMACRO{\QDATOP{\QDATOP{}{}}{\QDATOP{}{}}}%
%BeginExpansion
\genfrac{}{}{0pt}{0}{\genfrac{}{}{0pt}{0}{}{}}{\genfrac{}{}{0pt}{0}{}{}}%
%EndExpansion
t_{2}%
%TCIMACRO{\QDATOP{\QDATOP{}{}}{\QDATOP{}{}}}%
%BeginExpansion
\genfrac{}{}{0pt}{0}{\genfrac{}{}{0pt}{0}{}{}}{\genfrac{}{}{0pt}{0}{}{}}%
%EndExpansion
$ & $-\dfrac{iw}{iw+i}$ & $\dfrac{g}{g-1}$ & $\dfrac{\alpha}{g-1}$\\\hline
$%
%TCIMACRO{\QDATOP{\QDATOP{}{}}{\QDATOP{}{}}}%
%BeginExpansion
\genfrac{}{}{0pt}{0}{\genfrac{}{}{0pt}{0}{}{}}{\genfrac{}{}{0pt}{0}{}{}}%
%EndExpansion
t_{3}%
%TCIMACRO{\QDATOP{\QDATOP{}{}}{\QDATOP{}{}}}%
%BeginExpansion
\genfrac{}{}{0pt}{0}{\genfrac{}{}{0pt}{0}{}{}}{\genfrac{}{}{0pt}{0}{}{}}%
%EndExpansion
$ & $-\dfrac{w+1}{w}$ & $\dfrac{g-1}{g}$ & $-\dfrac{\alpha}{g}$\\\hline
$%
%TCIMACRO{\QDATOP{\QDATOP{}{}}{\QDATOP{}{}}}%
%BeginExpansion
\genfrac{}{}{0pt}{0}{\genfrac{}{}{0pt}{0}{}{}}{\genfrac{}{}{0pt}{0}{}{}}%
%EndExpansion
t_{4}%
%TCIMACRO{\QDATOP{\QDATOP{}{}}{\QDATOP{}{}}}%
%BeginExpansion
\genfrac{}{}{0pt}{0}{\genfrac{}{}{0pt}{0}{}{}}{\genfrac{}{}{0pt}{0}{}{}}%
%EndExpansion
$ & $-\dfrac{1}{w+1}$ & $\dfrac{1}{1-g}$ & $-\dfrac{\alpha}{1-g}$\\\hline
$%
%TCIMACRO{\QDATOP{\QDATOP{}{}}{\QDATOP{}{}}}%
%BeginExpansion
\genfrac{}{}{0pt}{0}{\genfrac{}{}{0pt}{0}{}{}}{\genfrac{}{}{0pt}{0}{}{}}%
%EndExpansion
t_{5}%
%TCIMACRO{\QDATOP{\QDATOP{}{}}{\QDATOP{}{}}}%
%BeginExpansion
\genfrac{}{}{0pt}{0}{\genfrac{}{}{0pt}{0}{}{}}{\genfrac{}{}{0pt}{0}{}{}}%
%EndExpansion
$ & $-\dfrac{iw+i}{i}$ & $1-g$ & $\alpha$%
\end{tabular}
\]

{\normalsize The first transformation has also been remarked and derived in a
different way in \cite{5}. The others are new. Note that in some
transformations (}$t_{2}${\normalsize \ and }$t_{5}${\normalsize ) the
coefficients must be imaginary to comply with the requirement of unit
determinant \textit{ad}-\textit{cb}=1.}

{\normalsize These transformations deserve the name of dualities because they
extend the theory from weak values of the statistical parameter }$g\in\left[
0,1\right]  ${\normalsize \ to strong values of the parameter }$g\in\left[
1,\infty\right]  ${\normalsize \ or the other way around. Indeed, starting
with a weak }$g\in\left[  0,1\right]  ${\normalsize , transformations }$t_{1}$
{\normalsize and }$t_{4}${\normalsize \ lead to formulas valid on the whole
real axis for \textit{g} and on the range }$g\in\left[  1,\infty\right]
${\normalsize , respectively. Starting with a strong }$g\in\left[
1,\infty\right]  ${\normalsize , transformation }$t_{3}${\normalsize \ leads
to formulas valid on the range }$g\in\left[  0,1\right]  ${\normalsize .
Transformations }$t_{2}${\normalsize \ and }$t_{5}${\normalsize \ map the
domain of validity for \textit{g} (both \textit{g} and }$\widetilde{g}%
${\normalsize \ must be necessarily positive) into themselves. The dual
distributions hold true at a temperature scaled with a function of \textit{g},
at energies below ( }$t_{2}${\normalsize \ and }$t_{5}${\normalsize , }%
$\alpha\geq0${\normalsize ) or above Fermi level (}$t_{1}${\normalsize ,}%
$t_{2}${\normalsize \ and }$t_{4}${\normalsize , }$\alpha\leq0${\normalsize ).}

{\normalsize A distribution with a greater than unit g supports two interpretations:}

({\normalsize i) as a distribution for particles; a physical realization is
the Hubbard model \cite{9}, where due to the infinite repulsion, two sites can
be occupied by at most one electron.(}$g=2${\normalsize ): and}

({\normalsize ii) as a distribution for vacancies. For instance, the
generalized exclusion principle (16) states that }$\frac{1}{g}$%
{\normalsize \ is the maximum number of particles a quantum state can
accommodate; then }$\frac{1}{\widetilde{g}}${\normalsize \ , where
}$\widetilde{g}${\normalsize \ is the dual of \textit{g} under }$t_{1}%
${\normalsize \ would represent the minimum number of states available for one particle.}

{\normalsize The relationships between a specific distribution and its dual
are given below}%

\begin{equation}
\widetilde{g}\cdot n\left(  \widetilde{\alpha},\widetilde{g}\right)  +g\cdot
n\left(  \alpha,g\right)  =1; \tag*{(20.1)}%
\end{equation}%

\begin{equation}
\left(  \widetilde{g}-1\right)  \cdot n\left(  \widetilde{\alpha}%
,\widetilde{g}\right)  +\left(  g-1\right)  \cdot n\left(  \alpha,g\right)
=1; \tag*{(20.2)}%
\end{equation}%

\begin{equation}
\left(  \widetilde{g}-1\right)  \cdot n\left(  \widetilde{\alpha}%
,\widetilde{g}\right)  +g\cdot n\left(  \alpha,g\right)  =1; \tag*{(20.3)}%
\end{equation}%

\begin{equation}
\widetilde{g}\cdot n\left(  \widetilde{\alpha},\widetilde{g}\right)  +\left(
g-1\right)  \cdot n\left(  \alpha,g\right)  =1; \tag*{(20.4)}%
\end{equation}%

\begin{equation}
n\left(  \widetilde{\alpha},\widetilde{g}\right)  +n\left(  \alpha,g\right)
=0; \tag*{(20.5)}%
\end{equation}

{\normalsize The last relationship is somehow peculiar. Apparently, due to
}$g\longrightarrow1-g${\normalsize \ transformation, it seems to convert the
distribution for fermions or near fermions }$\left(  g\geq\frac{1}{2}\right)
${\normalsize \ into the distribution for bosons or near bosons }$\left(
g\leq\frac{1}{2}\right)  ${\normalsize \ and vice-versa, but the dual
}$n\left(  \widetilde{\alpha},\widetilde{g}\right)  ${\normalsize \ of
}$n\left(  \alpha,g\right)  ${\normalsize \ is, in this case, negative}

{\normalsize The above dualities impose constraints on the expansion
coefficients of thermodynamic functions. For instance, the duality (20a) leads
to a vanishing zero temperature heat capacity (Refs. 5-8). A detailed
derivationand analysis of these constraints are relegated to an extended
paper. Here we are content to remark that all five transformations are needed
for these transformations to close upon themselves, i.e. to form a group. We
think the group property is the most remarkable feature and we give below the
table for the group composition law:}%

\[
\text{%
\begin{tabular}
[c]{c||c|c|c|c|c|c}%
\hspace{1cm}$%
%TCIMACRO{\QDATOP{{}}{{}}}%
%BeginExpansion
\genfrac{}{}{0pt}{0}{{}}{{}}%
%EndExpansion
$ & $\hspace{0.25cm}t_{0}\hspace{0.25cm}$ & $\hspace{0.25cm}t_{1}%
\hspace{0.25cm}$ & $\hspace{0.25cm}t_{2}\hspace{0.25cm}$ & $\hspace
{0.25cm}t_{3}\hspace{0.25cm}$ & $\hspace{0.25cm}t_{4}\hspace{0.25cm}$ &
$\hspace{0.25cm}t_{5}\hspace{0.25cm}$\\\hline\hline
$%
%TCIMACRO{\QDATOP{{}}{{}}}%
%BeginExpansion
\genfrac{}{}{0pt}{0}{{}}{{}}%
%EndExpansion
t_{0}%
%TCIMACRO{\QDATOP{{}}{{}}}%
%BeginExpansion
\genfrac{}{}{0pt}{0}{{}}{{}}%
%EndExpansion
$ & $t_{0}$ & $t_{1}$ & $t_{2}$ & $t_{3}$ & $t_{4}$ & $t_{5}$\\\hline
$%
%TCIMACRO{\QDATOP{{}}{{}}}%
%BeginExpansion
\genfrac{}{}{0pt}{0}{{}}{{}}%
%EndExpansion
t_{1}%
%TCIMACRO{\QDATOP{{}}{{}}}%
%BeginExpansion
\genfrac{}{}{0pt}{0}{{}}{{}}%
%EndExpansion
$ & $t_{1}$ & $t_{0}$ & $t_{3}$ & $t_{2}$ & $t_{5}$ & $t_{4}$\\\hline
$%
%TCIMACRO{\QDATOP{{}}{{}}}%
%BeginExpansion
\genfrac{}{}{0pt}{0}{{}}{{}}%
%EndExpansion
t_{2}%
%TCIMACRO{\QDATOP{{}}{{}}}%
%BeginExpansion
\genfrac{}{}{0pt}{0}{{}}{{}}%
%EndExpansion
$ & $t_{2}$ & $t_{4}$ & $t_{0}$ & $t_{5}$ & $t_{1}$ & $t_{3}$\\\hline
$%
%TCIMACRO{\QDATOP{{}}{{}}}%
%BeginExpansion
\genfrac{}{}{0pt}{0}{{}}{{}}%
%EndExpansion
t_{3}%
%TCIMACRO{\QDATOP{{}}{{}}}%
%BeginExpansion
\genfrac{}{}{0pt}{0}{{}}{{}}%
%EndExpansion
$ & $t_{3}$ & $t_{5}$ & $t_{1}$ & $t_{4}$ & $t_{0}$ & $t_{2}$\\\hline
$%
%TCIMACRO{\QDATOP{{}}{{}}}%
%BeginExpansion
\genfrac{}{}{0pt}{0}{{}}{{}}%
%EndExpansion
t_{4}%
%TCIMACRO{\QDATOP{{}}{{}}}%
%BeginExpansion
\genfrac{}{}{0pt}{0}{{}}{{}}%
%EndExpansion
$ & $t_{4}$ & $t_{2}$ & $t_{5}$ & $t_{0}$ & $t_{3}$ & $t_{1}$\\\hline
$%
%TCIMACRO{\QDATOP{{}}{{}}}%
%BeginExpansion
\genfrac{}{}{0pt}{0}{{}}{{}}%
%EndExpansion
t_{5}%
%TCIMACRO{\QDATOP{{}}{{}}}%
%BeginExpansion
\genfrac{}{}{0pt}{0}{{}}{{}}%
%EndExpansion
$ & $t_{5}$ & $t_{3}$ & $t_{4}$ & $t_{1}$ & $t_{2}$ & $t_{0}$%
\end{tabular}
}%
\]
{\normalsize where }$t_{0}${\normalsize \ denotes the identity. The left
multiplication corresponds to entries along column, the right multiplication
to entries along rows. It is worthy to note that transformations }$t_{j}%
${\normalsize \ , }$j=3,4${\normalsize \ are inverse of each other, a fact
that can also be seen from rels. (20.3), (20.4)}

\section{Conclusions}

{\normalsize To sum up, we have proved that the fractional statistics
distribution is invariant to a finite, nonabelian subgroup of the fractional
linear group.}

{\normalsize We have discussed the duality in terms of transformations that,
while leaving the theory invariant, relate two different ``coupling constant''
\textit{g} regimes. There exists, also, another meaning of duality in this
letter, as illustrated, for instance, in relation (4), namely, the
wave-particle duality. The first term in the RHS of rel. (4) features the
particle-like character of bosons and the second, the wave-like character. The
fact the two terms simply add togather expresses the complementarity of the
two features. In standard quantum mechanics, this duality is embodied in the
usual momentum-position commutator. We may wonder if the newly introduced term
}$g\left(  g-1\right)  n^{3}${\normalsize \ for fractional statistics, in
relation (12) signals the existence, in an analogous way, of another level of duality.}

{\normalsize In view of the above remarks it is tempting to put forward the
following conjecture: there must exist another type of duality, in the sense
above, but between vortices and particles (as in some models for anyons [11]).
The conjecture is rather speculative at the moment. We hope to return with a
more rigorous statement in the next future.}

\bigskip\textbf{Acknoledgement}

The autor {\normalsize thanks Gr. Ghika for critical reading of the
paper.\bigskip}

\end{document}